# VARIATIONS OF THE SELECTIVE EXTINCTION ACROSS THE GALACTIC BULGE — IMPLICATIONS FOR THE GALACTIC BAR

P. R. Woźniak[1]

Warsaw University Observatory, Al. Ujazdowskie 4, 00–478 Warszawa, Poland

e-mail: wozniak@vela.astrouw.edu.pl

K. Z. Stanek[2]

Princeton University Observatory, Princeton, NJ 08544–1001

e-mail: stanek@astro.princeton.edu

## ABSTRACT

We propose a new method to investigate the coefficient of the selective extinction, based on two band photometry. This method uses red clump stars as a means to construct the reddening curve. We apply this method to the OGLE color-magnitude diagrams to investigate the variations of the selective extinction towards various parts of the Galactic bulge. We find that $A_V/E_{V-I}$ coefficient is within the errors the same for $l = \pm 5 \deg$ OGLE fields. Therefore, the difference of $\sim 0.37$ $mag$ in the extinction adjusted apparent magnitude of the red clump stars in these fields (Stanek et al. 1994, 1995) cannot be assigned to a large-scale gradient of the selective extinction coefficient. This strengthens the implication of this difference as indicator of the presence of the bar in our Galaxy. However using present data we cannot entirely exclude the possibility of $\sim 0.2$ $mag/mag$ variations of the selective extinction coefficient on the large scales across the bulge.

*Subject headings:* Galaxy: general – Galaxy: structure – stars: Hertzsprung-Russell diagram – stars: statistics

---

[1]Visiting Student, Princeton University Observatory, Princeton, NJ 08544-1001

[2]On leave from N. Copernicus Astronomical Center, Bartycka 18, Warszawa 00–716, Poland

## 1. INTRODUCTION

Stanek et al. (1994, 1995) used the color-magnitude diagrams (CMDs) obtained by the OGLE experiment (Udalski et al. 1993, 1994 and references therein) to investigate the presence of the bar in our Galaxy. They found a many sigma difference of $\sim 0.37\ mag$ in the extinction-adjusted apparent magnitude of the red clump stars in the two opposite $l = \pm 5\deg$ OGLE Galactic bar fields. In the simplest scenario, this difference is a strong argument for our Galaxy being a barred galaxy. However, there remains a possibility that the observed magnitude offset between the two sides to the Galactic bulge could be produced by a large-scale gradient in the properties of the interstellar dust. This could cause the coefficient of the selective extinction for the $V$ and $I$ filters to be larger on one side of the Galactic bulge than on the other side (see discussion in Stanek et al. 1994).

In this paper we propose a new method for investigating the value of the coefficient of the selective extinction. This method uses red clump stars as a means to construct the reddening curve. We apply this method to the OGLE CMDs to investigate the value of the selective extinction coefficient for various fields across the Galactic bulge. Since in any given field there is a range of interstellar extinction, we can separate subfields with different extinction and construct a reddening curve. A similar approach was tried by Blanco et al. (1984), who divided their photographic plate of Baade's Window into four subfields according to the surface density of stars. However, their division was done in rather arbitrary fashion by visual inspection of the plate. The method we propose in this paper is entirely quantitative. In Section 2 we discuss the data used in this paper and the method for statistically separating subfields with different extinction. In Section 3 we use this method to construct the reddening curves for various fields across the Galactic bulge. In Section 4 we apply different checks to our results and we discuss the implications of these results for the Galactic bar investigated by Stanek et al. (1994, 1995).

## 2. THE DATA

We use two somewhat separate data products created by the OGLE collaboration. Szymański & Udalski (1993) constructed a database of photometric measurements in $V$ and $I$ bands for the fields observed by the OGLE project. Udalski et al. (1993) present color-magnitude diagrams (CMDs) of 13 fields in the direction of the Galactic bulge, which cover nearly one square degree and contain about $5 \times 10^5$ stars. All observations were made using the 1 meter Swope telescope at the Las Campanas Observatory, operated by the Carnegie Institution of Washington, and a $2048 \times 2048$ pixel Ford/Loral CCD detector with the pixel size 0.44 arcsec covering $15' \times 15'$ field of view. CMDs for all the fields analyzed in this paper can be seen in Udalski et al. (1993). Most of each diagram is dominated by bulge stars, with distinct red clump, red giant and turn-off point stars. The part of the diagram dominated by the disk stars was analyzed by Paczyński et al. (1994).

From the database (Szymański & Udalski 1993) we extract for each field a list of stars in this field along with their $V$ magnitude and the position on the frame. We decided to use only the stars with $V < 20.0$ $mag$, because the completeness of the fainter stars' detection in each frame depends on how crowded the field is. To distinguish between subfields of different extinction we divide each frame into $32 \times 32$ subframes and compute total number of database stars with $V < 20.0$ $mag$ in each subframe. There were on average between 15 and 100 stars in each subframe, with small variations for various fields. We then divide the subframes into 16 groups in order of increasing number of database stars per subframe. We make the division such that each group contains roughly the same total number of the red clump stars (as defined by our cut, see the next section). Assuming that the extinction affects significantly the number of stars detected in $V$ band in a given subframe, this ordering should reflect a decreasing extinction in a group. Then we add CMD stars belonging to each group. To illustrate how the described above procedure works, we show in Fig.1 region of the CMD for the BW3 field dominated by red clump stars. Only two groups, one with the highest (filled circles) and one with the lowest (open circles) density of database stars, are shown. The shift on the CMD is evident.

## 3. THE ANALYSIS

In the previous section we showed that the simple method we use for separating parts of the field with different extinction is very efficient (Fig.1). Now we want to utilize this method to obtain the quantitative values of the offset on the CMD between the different subfields caused by different extinction. We proceed as follows: using the method described in the previous section we divide the CMD for a given field into 16 separate groups. Then we use the red clump dominated parts of CMDs for determining the offsets. A suitable window in the CMD plane is described by the following inequalities:

$$1.4 < (V - I) < 2.6 \; ; \quad 11.0 < V - 2.6\,(V - I) < 13.5. \tag{1}$$

The factor of 2.6 in the above equation comes from using the reddening law $E_{V-I} = A_V/2.6$, following Dean, Warren, & Cousins (1978) and Walker (1985).

We take the fifth group as the reference group (it roughly falls in the middle of the extinction range) and compute the shifts between each of 16 distributions and the reference distribution. As a measure of the difference between two given distributions we adopted the Kolmogorov-Smirnov test. The shifts in $(V - I)$ and $V$ are computed separately using the same method, that is shifting one of the distributions to minimize corresponding Kolmogorov-Smirnov statistic $D$. We found that the shifts determined separately correspond exactly to the shifts found using a 2-D modification of the KS statistic (Press et al. 1992), but are much more CPU efficient.

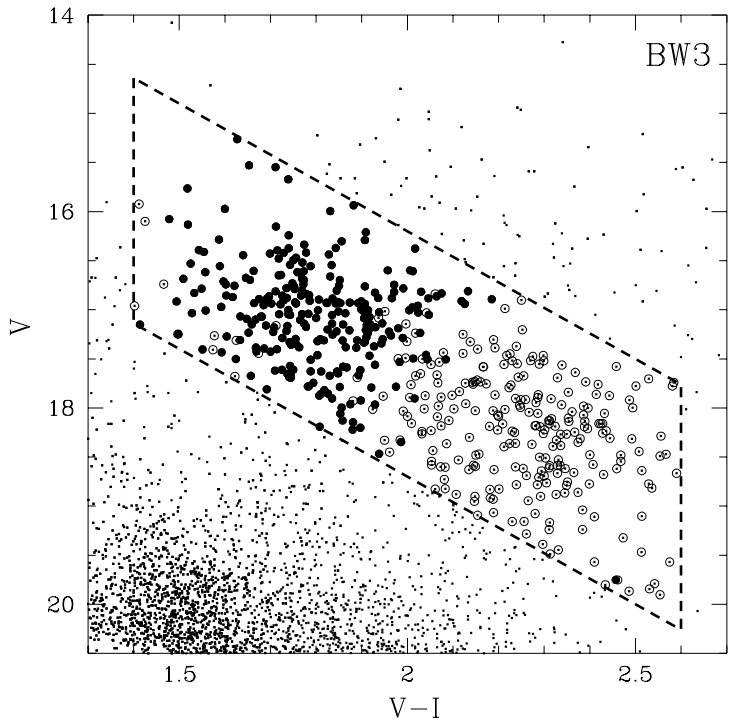

Fig. 1.— Part of the color-magnitude diagram dominated by red clump stars for the BW3 field. Shown with different symbols are stars from parts of the field with lowest (open circles) and highest (filled circles) extinction. This shows that the simple method, described in the text, for separating the parts of the field with different extinction, is very efficient. We also show, with the thick dashed line, the part of the CMDs which was selected for quantitative analysis of the reddening shift.

We thus obtained the shifts in reddening and extinction for each subfield, a point in the $\delta(V-I), \delta V$ plane. The next step was estimation of the random errors of these values. This was done using bootstrap technique. From the observed distribution of the red clumps stars we randomly created a new distribution, with the same number of stars, to which we then applied the described above method to find the offset in the $\delta(V-I), \delta V$ plane. By repeating this procedure 1000 times we obtained the probability distribution of each offset, which appeared to be very well represented by a gaussian. The centered ranges of $\delta(V-I)$ and $\delta V$ independently containing 68% of the stars in each set were taken to be $1\sigma$ errors for our values of the shifts in the reddening and the extinction. Once completed we had 16 points with errors in $\delta(V-I)$ and $\delta V$ for each field. The last step was the least-squares straight line fit to these points. To obtain the slope and its standard error we used the fitting procedure with errors in both axes described by Press et al. (1992). Such computed error of the slope was compatible with the dispersion of the gaussian distribution of slopes when fitted to each of 1000 bootstrap realizations separately. The example of our computations is shown in Fig.2.

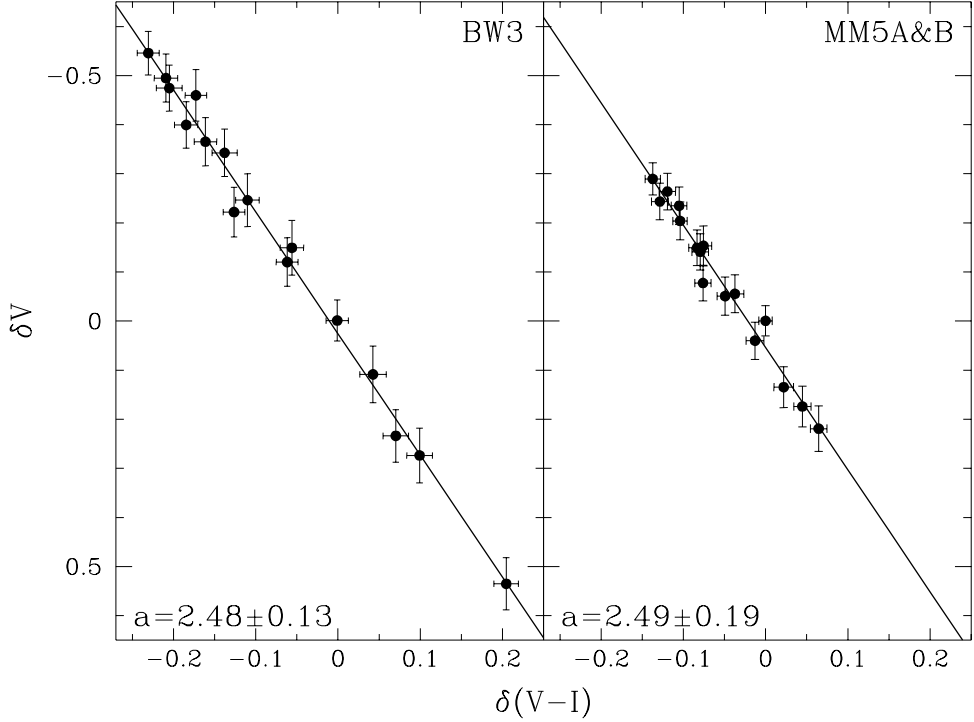

Fig. 2.— Reddening curve for the BW3 (left) and MM5 (right) fields. Each dot with the error bars corresponds to the offset and its $1\sigma$ errors resulting from bootstrap. Straight, continuous line represents the least-squares fit to the reddening curve using the resulting offsets and their errors.

We found that the range of extinction in the field had very strong influence on the final error of the slope determination. It is important to constrain the difference in the coefficient between MM5 and MM7 fields, so we decided to process MM5A and MM5B as one field to increase the range of extinction values covered with the one field and improve the statistics. The same was done for MM7 and four pairs of the BW fields. The choice for BW fields was made so that the fields processed together were adjacent. Final values of the reddening coefficient, its error and the range of the extinction within the field ($\Delta V = \delta V_{max} - \delta V_{min}$) are reported in Table 1 and shown in Fig.3. The weighted mean of these values is 2.44, somewhat smaller than the value 2.6 of Dean, Warren, & Cousins (1978) and Walker (1985). This agrees with the value of this coefficient found recently by Ng et al. (1995), using different methods.

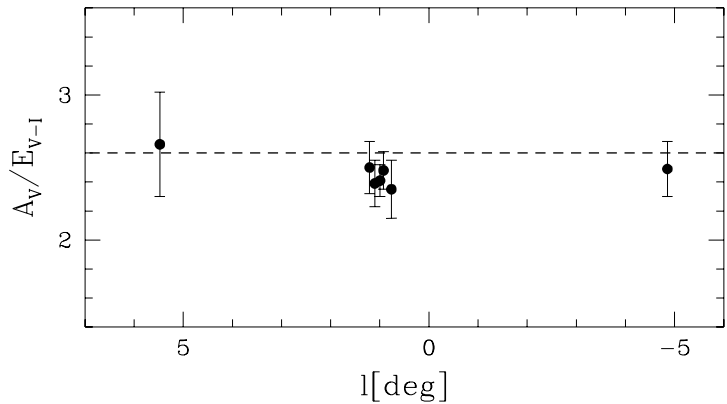

Fig. 3.— Value of the selective extinction coefficient plotted as a function of the Galactic longitude $l$. There is no obvious correlation between the two variables. The dashed line corresponds to the reddening law of Dean, Warren, & Cousins (1978) and Walker (1985).

## 4. DISCUSSION

In the previous section we obtained values and estimated random errors for the selective extinction coefficient for BW3 field and six pairs of BW and MM OGLE fields. However, there is a possibility of systematic errors in our determination of the reddening shift between different subfields. For example, the cut we applied to the CMDs in order to select the red clump regions (Fig.1) may cause systematic error.

We searched for possible bias in the offset values found by the algorithm we applied for determining the shifts in the CMD plane. We took our 16 CMD distributions for the BW3 field and corrected them using the shifts we obtained before. Thus all the distributions were roughly in the same reference frame of the fifth distribution. Then we constructed an artificially reddened CMD by shifting them again along the straight line of the known slope changing randomly their ordering, so that the shifts covered uniformly certain range $\Delta V$. To such constructed artificial CMD we applied our algorithm for finding $A_V/E_{V-I}$ to see if there are any systematic errors. This was repeated for 6 values of the slope ranging from 2.1 to 3.1 and 110 separations in the range $0.1 - 1.2\ mag$. The result of these simulations is shown in Fig.4. The diagram was smoothed for display purposes, so each point represents the mean value of the reproduced slopes in given $0.1\ mag$ range of separations. The solid line corresponds to the slope of the artificial reddening curve. It can be seen that the value of 2.6 adopted in our definition of the red clump is being preferred by the algorithm and that the error is $< 0.1\ mag/mag$ in the range of interest. It also shows that the range of extinction in the field is crucial for determination of the selective extinction coefficient. Although our procedure is not perfect, we decided not to correct for the systematic shifts

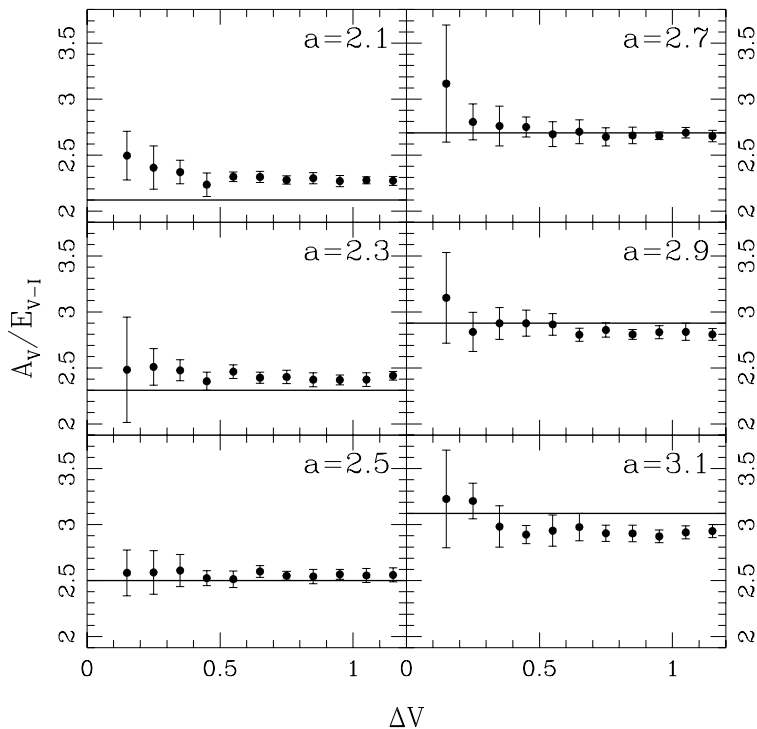

Fig. 4.— Systematic errors. Each point with the error bar corresponds to ten artificial reddening curves and represents the mean of their slopes and its standard deviation. The solid line corresponds to the value of the slope that was actually used to construct an artificial reddening curve.

as they are not large compared to the random errors and it would have not changed our conclusions.

As an additional check, we recomputed the values of $A_V/E_{V-I}$ using $I$ database counts as a basis for separating subfields with different extinction. The resulting coefficients were consistent with the ones obtained using $V$ counts, however they had slightly larger errors.

It is tempting to see whether the method we described in this letter can be used to correct the CMD for the differences in the selective extinction within the field. The offsets for the BW3 field, resulting from our algorithm, were applied to the magnitudes and colors for each of the 16 parts of the CMD for this field. The result is the CMD in the reference frame of the comparison distribution (see the previous section), which has intermediate extinction. A corrected CMD for the BW3 field, as well as a version with no correction applied, are displayed in Fig.5. The dereddened CMD is less diffuse, shows much more compact red clump and a narrower band of red giants.

As seen in Table 1, the results obtained by us are consistent with a constant coefficient of selective extinction throughout the Galactic bulge. However, with the existing data we cannot entirely exclude the possibility of small ($\sim 0.2\ mag/mag$) variations of this coefficient on the large scale. The main reason is that the OGLE fields we analyzed in this paper were originally chosen to have *small* extinction and are therefore not ideally suited for our purpose. As it happens both MM5 and MM7 fields have small variation of the extinction in the field, which is confirmed by visual inspection of the Palomar Survey plates. We plan to obtain photometry for additional fields adjacent to the existing ones, but with much higher extinction. This should allow us to obtain a much better constraints on the variation of the reddening law across the Galactic bulge. We can however already say that it is very unlikely that the difference of $\sim 0.37\ mag$ in the extinction-adjusted apparent magnitude of the red clump stars in the MM5 and MM7 fields (Stanek et al. 1994) is produced entirely by the variations in the reddening law.

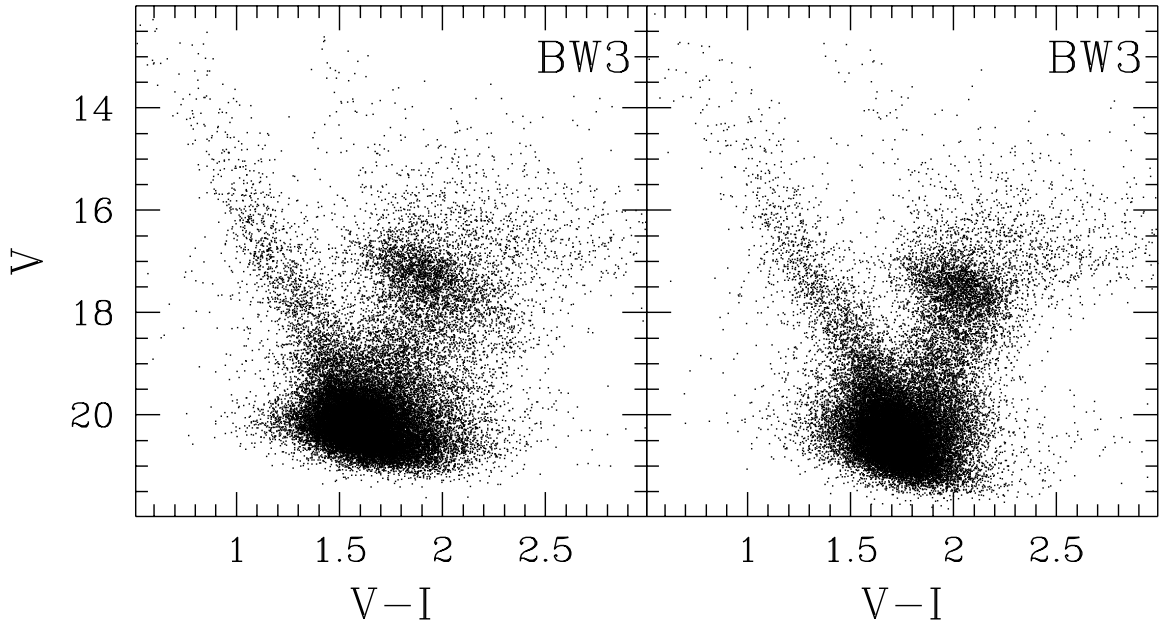

Fig. 5.— CMD for the BW3 field as from OGLE data with no corrections (left panel) and corrected for the differences of the selective extinction within the field (right panel).

We would like to thank the OGLE collaboration for making their data available to us. We also thank Bohdan Paczyński for constant support and many stimulating discussions. Wes Colley provided us with useful comments on an earlier version of this paper. This project was supported with the NSF grant AST 9216494. KZS thanks also for the NAS Grant-in-Aid of Research through Sigma Xi, The Scientific Research Society.

# REFERENCES

Blanco, V. M., McCarthy, M. F., & Blanco, B. M., 1984, AJ, 89, 636

Dean, J. F., Warren, P. R., & Cousins, A. W. J., 1978, MNRAS, 183, 569

Ng, Y. K., Bertelli, G., Chiosi, C., & Bressan, A., 1995, A&A, submitted

Paczyński, B., Stanek, K. Z., Udalski, A., Szymański, M., Kałużny, J., Kubiak, M., & Mateo, M. 1994, AJ, 107, 2060

Press, W. H., Flannery, B. P., Teukolsky, A. S., & Vetterling , W. T. 1992, Numerical Recipes (Cambridge: Cambridge Univ. Press)

Stanek, K. Z., Mateo, M., Udalski, A., Szymański, M., Kałużny, J., Kubiak, M., 1994, ApJ, 429, L73

Stanek, K. Z., Mateo, M., Udalski, A., Szymański, M., Kałużny, J., Kubiak, M., & Krzemiński, W., 1995, in: IAU Symposium 169, "Unsolved Problems of the Milky Way", ed. L. Blitz, in press

Szymański, M., & Udalski, A. 1993, Acta Astron., 43, 91

Udalski, A., Szymański, M., Kałużny, J., Kubiak, M., & Mateo, M. 1993, Acta Astron., 43, 69

Udalski, A., Szymański, M., Stanek, K. Z., Kałużny, J., Kubiak, M., Mateo, M., Krzemiński, W., Paczyński, B., & Venkat, R. 1994, Acta Astron., 44, 165

Walker, A. R., 1985, MNRAS, 213, 889

Table 1. Selective extinction coefficients.

| Field | $l$ [deg] | $A_V/E_{V-I}$ [$mag/mag$] | $\Delta V$ [$mag$] |
|---|---|---|---|
| BW3 | 0.923 | $2.48 \pm 0.13$ | 1.08 |
| BW1 + BW5 | 0.993 | $2.41 \pm 0.11$ | 0.92 |
| BW2 + BW6 | 0.764 | $2.35 \pm 0.20$ | 0.56 |
| BW4 + BW7 | 1.208 | $2.50 \pm 0.18$ | 0.82 |
| BW8 + BWC | 1.102 | $2.39 \pm 0.16$ | 0.45 |
| MM5A + MM5B | $-4.853$ | $2.49 \pm 0.19$ | 0.51 |
| MM7A + MM7B | 5.478 | $2.66 \pm 0.36$ | 0.45 |